\documentclass[letterpaper, 10 pt, conference]{ieeeconf}  

\IEEEoverridecommandlockouts                              

\usepackage{bbold}
\usepackage{graphics} 
\usepackage{graphicx}
\usepackage{epsfig} 
\usepackage{times} 
\usepackage{balance}
\usepackage{color}
\usepackage{epstopdf}
\usepackage[utf8]{inputenc}

\usepackage{amsmath}
\usepackage{amssymb}
\usepackage[colorlinks,
linkcolor=black,
anchorcolor=black,
citecolor=black
]{hyperref}

\usepackage{subcaption}

\usepackage{booktabs}
\usepackage{mathrsfs}
\usepackage{theorem}
\usepackage{algorithm}
\usepackage{algorithmic}
\usepackage{tabularx}
\usepackage{indentfirst}
\usepackage{float}
\usepackage{stfloats}

\DeclareMathOperator*{\argmin}{arg\,min}

\newcommand{\tr}{\top} 

\title{\LARGE \bf  Online power system parameter estimation  and optimal operation
}

\author{Xu Du, Alexander Engelmann, Timm Faulwasser and Boris Houska  
	\thanks{XD and BH are with the School of Information Science and Technology, ShanghaiTech University, China. 
		{\tt \{duxu, borish\}@shanghaitech.edu.cn }
	}%
\thanks{TF and AE are with the Institute of Energy Systems, Energy Efficiency and Energy Economics, TU Dortmund University, Germany. 
	Main parts of this work have been conducted while AE was with 
	the Institute for Automation and applied Informatics, Karlsruhe Institue of
	Technology, Germany.
	{\tt alexander.engelmann@tu-dortmund.de, timm.faulwasser@ieee.org}}%
	\thanks{DX and BH are supported by ShanghaiTech University, Grant-Nr. F-0203-14-012.  }
}

\begin{document}
	
\maketitle
\thispagestyle{empty}
\pagestyle{empty}

\begin{abstract}
	The integration of renewables into electrical grids
	calls for novel control schemes, which usually are model based. Classically, for power systems parameter estimation and optimization-based control are often decoupled, which may lead to increased cost of system operation during the estimation procedures. The
	present work proposes a method for simultaneously minimizing grid operation cost and estimating line parameters. To this end, we rely on  methods from optimal design of experiments. This approach leads to a substantial reduction in cost for optimal estimation and in higher accuracy in the parameters compared
	with standard combination of optimal power flow and maximum-likelihood estimation. We illustrate the performance of the proposed method on simple benchmark system.

\end{abstract}

\section{Introduction}
Economical optimal and safe operation of power systems with a large share of renewables requires reliable grid models.
While the grid topology is often known, the parameters are are frequently unknown or erroneous \cite{Abur2004,Kusic2004}.
A classical approach for the optimal operation of power systems is to run an estimation procedure obtaining grid parameters first, and secondly using these parameters in an Optimal Power Flow~(OPF) problem for computing optimal generator set-points.
Although this procedure is usually reliable, it may lead to high system operation cost until the estimation is converge to accurate parameters.

Established theory of maximum-likelihood estimation and Bayesian methods can be found in many textbooks \cite{kay1993fundamentals,Ljung1999}.
For static power system parameter estimation based on multiple measurement snapshots,  recursive least-squares based techniques have been proposed in \cite{Bian2011,Slutsker1996,VanCutsem1988}.
Approaches for combined parameter and topology estimation are considered in \cite{Deka2016,Park2018}. 
These works do typically consider constant or given power injections.

Recently, parameter estimation based on techniques from Optimal Experiment Design (OED)  have been suggested \cite{Du2020, fabbiani2020identification}.\footnote{Consider \cite{pukelsheim2006optimal} for an excellent presentation of the foundations of OED.}
Therein, the conceptually new idea is to compute set points for generators such that a maximum amount of information is extracted in each time instant leading to a fast and accurate estimation. 
To this end, an optimization problem is constructed, which minimizes the trace of the  covariance matrix of the parameter estimates.
Although this procedure leads to fast and accurate estimates, it also induces to high cost of system operation in the estimation procedure since it is agnostic to the associated economic costs of choosing the set points optimally with respect to the estimation variance.

A second branch considers an \emph{economic} variant of OED by considering the cost of experiments in the design procedure \cite{Houska2015}. 
This approach is combined with model predictive control  \cite{Feng2018} and with power system parameter estimation \cite{Du2020}.
%

In the present work we propose a similar approach for power system parameter estimation aiming at lowering the operation cost in the estimation procedure.
We compute a Pareto front trading off system operation cost via OPF versus the goal of obtaining highly accurate grid parameters. 
Based on this curve, we develop a scheme for adjusting the weighting parameter in the estimation step to reach a predefined accuracy in the parameter estimates after a desired number of sampling instants.


The remainder of this paper is organized as follows: \autoref{sec:gridModel} recalls AC grid modeling basics. 
In \autoref{sec:OED-OPF} we describe the main contribution of this paper: a method trading-off estimation with optimal system operation based on optimal design of experiments. 
\autoref{sec:numRes} shows promising numerical results in terms of a  higher accuracy in the parameters compared with classical maximum-likelihood estimation methods and a substantially reduced operation cost compared to classical OED.

\section{The AC Grid Model and Optimal Power Flow} \label{sec:gridModel}
This section recalls basics of  AC power system modeling and the AC OPF problem, which serves as a basis for our developments.

\subsection{Power Grid Model}
We consider a power grid  $(\mathcal{N},\mathcal{L})$, where $\mathcal{N}=\{1\dots N\}$ denotes the set of buses and $\mathcal{L}\subseteq \mathcal{N}\times\mathcal{N}$ represents the set of transmission lines. 
The physical properties of transmission lines are described by line conductances $g_{k,l}$ and line susceptances $b_{k,l}$ for all transmission lines $(k,l) \in \mathcal L$, which we would like to estimate.
We set $g_{k,l} = b_{k,l} =0$ for all $(k,l) \notin \mathcal L$ \cite{Grainger1994,Frank2016}.
We collect these line parameters in 
\begin{equation}\label{eq:y}
	y = \left(
	\begin{array}{c}
		g_{k,l} \\[0.16cm]
		b_{k,l} 
	\end{array}
	\right)_{(k,l)\in \mathcal{L}} \in \mathbb{R}^{2|\mathcal{L}|} \notag,
\end{equation}
where $(v_k)_{k \in \mathcal{S}}$ denotes the vertical concatenation of vectors $v$ over the index set $\mathcal{S}$.

Denoting the voltage amplitude at node $k$ by $v_k$   and  the voltage magnitude at node $k$ by $\theta_{k}$, we define the algebraic state of the  system  as
\[
x = \left( v_2, \theta_2, v_3, \theta_3, \ldots, v_N, \theta_N \right)^\tr.
\]
The voltage magnitude and the voltage angle at the first node, the slack node, are assumed to be fixed and given
\[
\theta_1 = 0 \quad \text{and} \quad v_1 = \mathrm{const.} 
\]
The active and reactive power flow over a transmission line  $(k,l) \in \mathcal{L}$ is given by
\begin{multline*}\label{eq:pi}
\Pi_{k,l}(x,y) = v_k^2 \left(
\begin{array}{r}
g_{k,l} \\[0.16cm]
- b_{k,l} 
\end{array}
\right) \\- v_kv_l
\left(
\hspace{-0.05cm}
\begin{array}{rr}
g_{k,l}  & b_{k,l} \\[0.16cm]
-b_{k,l} & g_{k,l} 
\end{array}
\hspace{-0.05cm}
\right)
\hspace{-0.05cm}
\left(
\hspace{-0.05cm}
\begin{array}{c}
\cos(\theta_{k}-\theta_{l}) \\[0.16cm]
\sin(\theta_{k}-\theta_{l})
\end{array}
\hspace{-0.05cm}
\right).
\end{multline*}
Summing up the transmission line flows of all nodes neighbored to node $k$ yields the residual active and reactive power 
\begin{align*}
S_k(x,y) &= \sum_{l \in \mathcal N_k} \Pi_{k,l}(x,y).
\end{align*}

Let $p^\mathrm{g}_k$ and $q^\mathrm{g}_k$ denote the active and reactive power generation at node $k$.
Then we set  $p^g_k = 0$ and $q^g_k = 0$ for all $k \notin \mathcal G$, where   $\mathcal G \subseteq \mathcal N$ denotes the set of generators of the system. 
We denote the active power demand at node~$k$ by $p^d_k$ and the reactive power demand at node~$k$ by $q^d_k$ . 
Both  are assumed to be known and constant and we set $p^d_k = 0$ and $q^d_k = 0$ in case there is no consumer at node $k$. 
We assume that the active and reactive power generation of all generators except at the first node are the only values we can control.
Hence, we introduce an input vector $u$ and a vector of power demands $d$ as
\begin{align}
\notag
u= \left(
\begin{array}{c}
p_k^\mathrm{g} \\[0.16cm]
q_k^\mathrm{g}\\
\end{array}
\right)_{k \in \mathcal N \setminus \{ 1 \} } \;\; \text{and} \;\;\; 
d= \left(
\begin{array}{c}
p_k^\mathrm{d} \\[0.16cm]
q_k^\mathrm{d}\\
\end{array}
\right)_{k \in \mathcal N \setminus \{ 1 \} }. \; 
\end{align}
With that, the so-called AC power flow equations \cite{Frank2016} are given by
\begin{equation}\label{eq:power_flow}
S(x,y) = u-d ,
\end{equation}
where 
\begin{equation*}
S(x,u) = \big(S_k(x,u) \big )_{k \in \mathcal N \setminus \{ 1 \} }.
\end{equation*}

\subsection{Optimal Power Flow}
Optimal Power Flow  aims at minimizing the total cost of power generation in an electrical grid subject to the power flow equations and physical and technical limits such as voltage bounds, line limits, and generator limits \cite{Frank2016,Zhu2015},
\begin{equation} \label{eq:OPF}
	\begin{aligned} 
		\min_{x,u}&\;\;C(u) \quad\\ \;\;\text{s.t.}&\left\{
		\begin{array}{l}
		S(x,y) = {u}-d   \\
			H(x,u)\leq 0 
		\end{array}.
		\right.
	\end{aligned}
\end{equation}
Here, $S$ encodes the power flow equations~\eqref{eq:power_flow}.
The cost function is typically quadratic in the active power generation
\begin{equation}\label{eq:OPF2}
		C(u) = \sum_{i\in\mathcal{G}} \alpha_i(p_i^g)^2 +\beta_i(p_i^g) \;,\;\\
\end{equation}
where $\alpha_i$ and $\beta_i$ are given and positive cost coefficients. 
Other formulations are possible, for example aiming at minimizing grid losses \cite{Frank2016}.
Moreover, the limits on voltages and power generation are modelled as 
\begin{equation*}
	\begin{split} 
		&H(x,u)= \left(
		\begin{array}{ll}
			p_k^g- \bar{p}_k\\
			q_k^g- \bar{q}_k\\
			x_k - \bar{x}_k\\
			\underline{p}_k- p_k^g\\
			\underline{q}_k- q_k^g\\
			\underline{x}_k- x_k\\
		\end{array}
		\right)_{k \in \mathcal{N}}.
	\end{split}
\end{equation*}

\section{Optimal Experiment Design for OPF} \label{sec:OED-OPF}
Next, we develop a method, which simultaneously estimates grid parameters and computes optimal generator set points. This leads to a higher accuracy in parameter estimates and lower operation cost compared to classical methods. 


\subsection{ Maximum Likelihood Parameter Estimation}
Maximum likelihood estimation based on least-squares techniques is a standard method for estimation of line parameters in power systems \cite{Slutsker1996,VanCutsem1988}.
The estimates typically rely on (noisy) measurements of transmission line flows, voltage magnitude and angle measurements at buses.
In this paper, we assume that all the active and reactive power flows through the transmission lines as well as the states of buses can be measured.
Thus, the measurement function $M$ is 
\[
M(x,y) = (x^\top, \Pi_{k,l}(x,y)^\top_{(k,l)\in\mathcal{L}})^\top.
\]
We assume additive Gaussian measurement noise with zero mean and a given variance $\Sigma\in \mathbb{S}^{|M|}_{++}$.

The corresponding Maximum Likelihood Estimation (MLE) problem for the line parameters $y$ is
\begin{equation}\label{eq:MLE}
\begin{aligned}
\min_{x\in\mathbb{R}^{2|\mathcal{N}|-2},y\in \mathbb{R}^{2|\mathcal{L}|} }&\;\; \frac{1}{2}\|M(x,y)-\eta\|_{\Sigma^{-1}}^2 + \frac{1}{2}\|y-\hat{y}\|^2_{\Sigma^{-1}_0}\quad\\\text{s.t.}&\left\{
\begin{array}{l}
S(x,y) = \hat{u}-d \\
H(x,u)\leq 0 
\end{array}
\right.
\end{aligned}.
\end{equation}
Here, $\hat{y}$ is the given initial parameter estimate with a given variance $\Sigma_0\in \mathbb{S}^{2|\mathcal{L}|\times 2|\mathcal{L}|}_{++}$ and $\eta$ are  measured values and $\hat{u}$ donates the current system input.

\subsection{Optimal Experiment Design}

Problem~\eqref{eq:MLE} depends the system inputs $u$.
Optimal experiment design exploits this degree of freedom by choosing the inputs such that a ``maximum amount of information'' is extracted.
The Fisher information matrix $\mathcal F \in \mathbb{S}^{2|\mathcal{L}|\times 2|\mathcal{L}|}_{++}$ of \eqref{eq:MLE} encodes the information content in the parameter estimates~\cite{Pukelsheim1993}  and is given by 
\begin{equation*}\label{eq:Fisher} 
	\begin{aligned}
	\mathcal{F}(x,u,\hat{y}) = \Sigma^{-1}_0 +  M_p(x,u,\hat{y})^{\top}\Sigma^{-1}M_p(x,u,\hat{y}).
	\end{aligned}
\end{equation*}
Here, we use the shorthand
\[M_p(x_s(u,\hat{y}),u,\hat{y})=-\frac{\partial M}{\partial x}\left(  \frac{\partial S}{\partial x}\right)^{-1}\frac{\partial S}{\partial p},
\]
where $ x_s(u,\hat{y})$ is the implicit solution of the equation $S(x,\hat y) = {u}-d$, which we assume to be unique \cite{Telen2013,Houska2015}.
Thus, the associated OED problem can be written as
\begin{equation} \label{eq:oed}
\begin{aligned} 
\min_{x,u}&\;\;\text{Tr}(\mathcal{F}(x,u,\hat{y})^{-1})+c\cdot\|p^g-p^{g-}\|^2_2\\ \;\;\text{s.t.}&\left\{
\begin{array}{l}
S(x,\hat y) = u-d \\
H(x,u)\leq 0,
\end{array}.
\right.
\end{aligned}
\end{equation}
where $p^{g-}$ denotes the previous system active power input. 

\subsection{Combining OPF and OED}

Problem~\eqref{eq:oed} computes optimal inputs $u$ such that the inverse of $\mathcal{F}$ (and thus the variance in the line parameters) is minimized. 
However, since  \eqref{eq:oed} is agnostic to the induced extra cost of this approach, we introduce a combined OPF-OED problem next, which simultaneously minimizes system operation cost and variance in the line parameters.

For safety reasons, it is often required that the line parameters are known up to a certain accuracy (e.g. to avoid line congestion).
Hence, one way to combine estimation with optimal operation is to pre-specify a certain target variance $\mathbb{V}_N^f$, which is necessary for safe operation and should be reached.
Such a target variance can typically not be reached in one step. 
Hence, one approach is to specify that after $N$ time steps, the target variance $\mathbb{V}_N^f$ should be reached and that this target variance should be reached as cheaply as possible.
In these time steps, some bounds might be violated. 
However, in many cases short-term overloading is possible due to thermal inertia of components. 
 
The above problem can be formulated as a multi-stage OED-OPF problem 
\begin{equation} \label{eq:multi-OED}
\begin{aligned} 
\min_{[x_k,u_k]}&\;\;\sum_{k=1}^{N}C(u_k)\\ \;\;\text{s.t.}&\left\{
\begin{array}{l}
\mathrm{Tr}(\mathbb{V}_N(x_1,u_1,y_1, \ldots, x_N,u_N,y_N))\leq\mathrm{Tr}(\mathbb{V}_N^f)\\
S(x_k,y) = u_k-d_k \\
H(x_k,u_k)\leq 0 ,
\end{array}
\right.
\end{aligned}
\end{equation}\vspace{-2pt}
where 
\begin{align*}
\begin{array}{l}
\mathbb{V}_N(x_1,u_1,y_1, \ldots, x_N,u_N,y_N) = \\[0.16cm]
\left (\Sigma^{-1}_0+\sum_{k=1}^{N}M_p(x_k,u_k,y_k)^{\top}\Sigma^{-1}M_p(x_k,u_k,y_k)\right )^{-1}
\end{array}
\end{align*}
is the predicted variance at time step $N$. Here, $[x_k,u_k]\in\mathbb{R}^{(4|\mathcal{N}|-4)\times|N|}$ indicate the decision variables of~\eqref{eq:multi-OED} for $N$ stages and $\Sigma^{-1}_0$ denotes the initial Fisher Information of the parameters.

\begin{algorithm}
	\small
	\caption{Autotuned OED-OPF tradeoff for parameter estimation}
	\textbf{Input:} Initial guess of parameter $\hat{y}$
	and variance $\mathbb{V}_0\succ 0$, an initial $\rho>0$, a termination tolerance $\epsilon > 0$, an initial generator set-point $\hat u$, and a terminal variance  $\mathrm{Tr}(\mathbb{V}_N^f)$.
	\\
	\textbf{Initialization:} $k=1$.\\
	\textbf{Repeat:}
	\begin{enumerate}
		
		\item  \emph{Collection of Measurements:} set the active and reactive power at the generators to u and take a measurement $\eta$.

		\item \emph{Maximum Likelihood Estimation:} Get new measurement $\eta$ and solve \textit{Estimation problem} (\ref{eq:MLE}) 
		\begin{equation*}
		\begin{split}
		(x_s,\hat{y}^+)=&\;\;\text{arg}\min_{x,y}\;\;  \frac{1}{2}\|M(x,y)-\eta\|_{\Sigma^{-1}}^2 + \frac{1}{2}\|y-\hat{y}\|^2_{\Sigma^{-1}_0}\quad\\
		&\text{s.t.}\;\;\left\{
		\begin{array}{l}
		S(x,y) = u-d \\
		H(x,u)\leq 0 
		\end{array}
		\right.
		\end{split}	
		\end{equation*}

		\item \emph{Set} $\mathbb{V}^+ =\mathrm{ Tr}(\mathcal{F}(x_s, \hat u,\hat y^+)^{-1})$.
		\vspace{0.1cm}
		\item  \emph{Update the mean of the expectation gap for the remaining steps} $I^+=\frac{1}{N-k}(\frac{1}{\mathrm{Tr}(\mathbb{V}_N^\mathrm{f})}-\frac{1}{\mathrm{Tr}(\mathbb{V}^+)})$.
		\vspace{0.1cm}
		\item \emph{Update weight $\rho$:} $\rho \leftarrow \rho+K(I^+-I_0)$.
		\vspace{0.1cm}
		\item \emph{Experiment Design:} Solve OED+OPF problem (\ref{eq:OED_OPF})	
		%
		and perform a new experiment with $u^*(\hat{y})$.
		
		\item \textit{Termination Criterion:} 
		If $\mathrm{Tr}(\mathbb{V}^+)<\epsilon$ for a small $\epsilon >0$, stop.
		
		\item \emph{Update:} Otherwise, set $\Sigma_0\leftarrow \mathbb{V}^+$, $u\leftarrow u^*(\hat y)$ and $\hat{y} \leftarrow \hat{y}^+$ and return to step 1) with $k\leftarrow k+1$.
	\end{enumerate}
	\label{alg:OED}
\end{algorithm}

In general it is hard to say whether the desired target variance is strictly reachable within $N$ steps since these predictions are supported by wrong parameters in each step. 
Moreover, such a multi-stage problem is also hard from a computational perspective---especially in case of large-scale grids. 
One way this issue is to relax the variance constraint to the objective function in a single-stage setting 
\begin{equation} \label{eq:OED_OPF}
\begin{aligned} 
(x^*,u^*)=\;\;\argmin_{x,u}\;\; 	&  C(u) + \frac{1}{\rho} \text{Tr}(\mathbb{V}(x,u,\hat{y})) \quad\\ \;\;\text{s.t.}&\left\{
\begin{array}{l}
S(x,\hat y) = u-d \\
H(x,u)\leq 0 
\end{array},
\right.
\end{aligned}
\end{equation}
where $\rho >0$ is penalty parameter.\footnote{Note that we solve \eqref{eq:OED_OPF} to \emph{local} optimality only. In the context of power systems, global optimality can usually not be guaranteed due to the non-convexity of the problem.}

\subsection{Adaptive Strategy for $\rho$}
\label{sec:adaptive}
\begin{figure*}[h]
	\centering
		\begin{subfigure}[t]{.49\textwidth}
		\centering
		\includegraphics[width=0.95\linewidth]{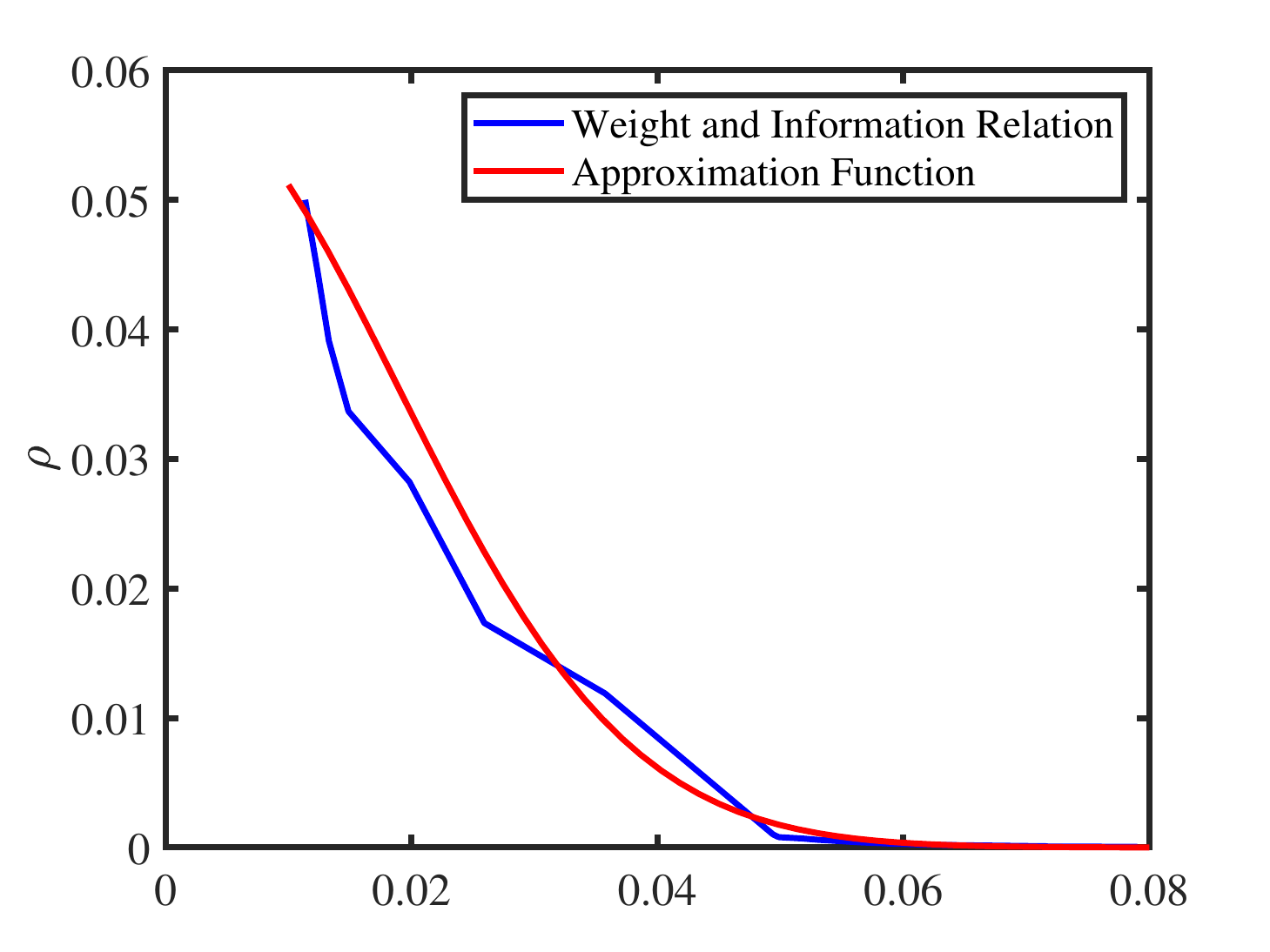}
		{\footnotesize{\vspace{-0.3cm}$$\frac{1}{\mathrm{Tr} \left\{ \mathbb{V}( x^\star(\rho,\hat y), u^\star(\rho,\hat y) , \hat y ) \right\} }\; [1/\mathrm{S}^2]$$}}
		\vspace{-0.25cm}
		\caption{Relationship between $\rho$ and  $\frac{1}{\mathrm{Tr}(\mathbb{V})}$.}
		\label{fig:V-rho}
	\end{subfigure}
	\hfill
		\begin{subfigure}[t]{.5\textwidth}
		\centering
		\includegraphics[width=0.95\linewidth]{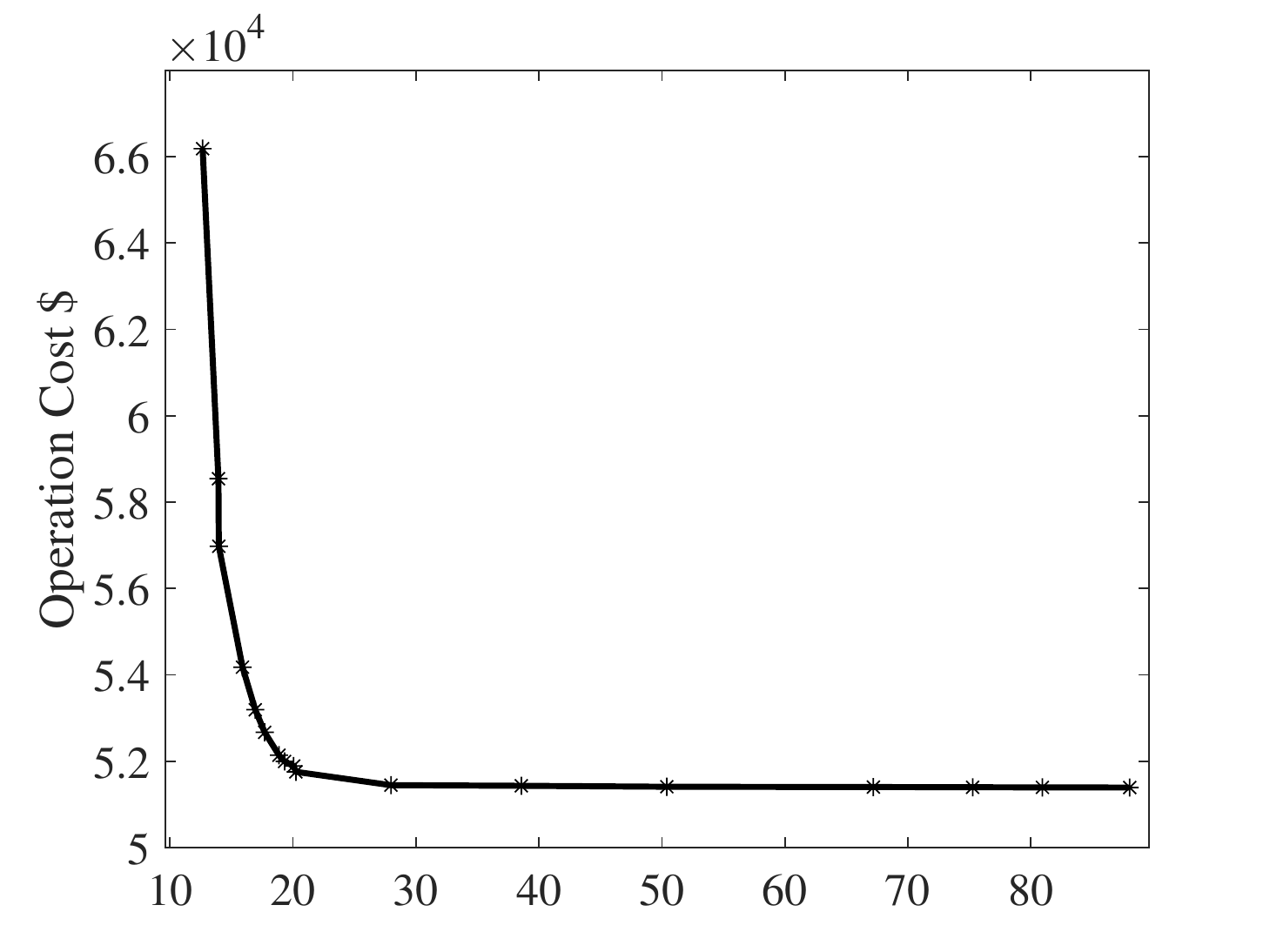}
		{\footnotesize{\vspace{-0.1cm}$${\mathrm{Tr} \left\{\mathbb{V}(x^\star(\rho,\hat y), u^\star(\rho,\hat y) , \hat y ) \right\} }\; [\mathrm{S}^2]$$}}
		\vspace{-0.35cm}
		\caption{Pareto optimality curve corresponding with OED and OPF}
		\label{fig:Pareto}
	\end{subfigure}
	\caption{Infuence of the weighting parameter $\rho$ on $\mathrm{ Tr}(\mathbb{V})$ and on the operation cost. }
\end{figure*}

We use an adaptive strategy for $\rho$ to reach our variance target.
Note that the target information we would like to have after $N$ iterations is $\frac{1}{\mathrm{Tr}(\mathbb{V}_N)}$.
Thus, the average information that we have to collect in each iteration is \vspace{-3pt}
\[
I_0=\frac{1}{N}\left (\frac{1}{\mathrm{Tr}(\mathbb{V}_N^f)}-\frac{1}{\mathrm{Tr}(\mathbb{V}_0)} \right ),
\] 
where $\mathrm{Tr}(\mathbb{V}_0)$ is the trace of initial parameter variance. 
The information to be gathered after iteration $k$ to the final iteration $N$ thus is \vspace{-3pt}
\[
I^+=\frac{1}{N-k}\left (\frac{1}{\mathrm{Tr}(\mathbb{V}_N^f)}-\frac{1}{\mathrm{Tr}(\mathbb{V}^+)}\right),
\] 
where $\mathbb{V}^+=\mathcal{F}(x_s,\hat{u},\hat{y})^{-1}$ is the realized variance after the estimation step \eqref{eq:MLE}, where $x_s$ denotes the  solution of \eqref{eq:MLE}.
With that, we introduce the update strategy
\begin{align} \label{eq:rhoUpdate}
\rho \leftarrow \rho+K(I^+)(I^+-I_0), 
\end{align}
tracking the average information we have to gather.
Here, $K(I^+)$ is an information state dependent feedback gain that is set to \vspace{-5pt}
\[
K(I^+) = \varphi'(I^+) \; ,
\] 
where $\varphi: \mathbb R \to \mathbb R$ is an approximation of the inverse trade-off function such that
\[
\varphi \left( \frac{1}{\mathrm{Tr} \left\{ \mathbb{V}( x^\star(\rho,\hat y), u^\star(\rho,\hat y) , \hat y ) \right\} } \right) \approx \rho \; .
\]
Notice that details on how to pre-compute approximations of this function as well the control gain $K(I^+)$ can be found in Section~\ref{sec:numRes}. In the above equation, $x^\star$ and $u^\star$ are the minimizers of~\eqref{eq:OED_OPF} depending on $\rho$ and $\hat y$.

The combined OED-OPF procedure is illustrated in Algorithm~\ref{alg:OED}. 
In the first step, the algorithm is initialized with trial values for the inputs $u^-$ and the line parameters $y^-$.
The guess for $u$ and $y$ are applied to the system and new measurements $\eta$ are arecollected.
These measurements are put in to the maximum-likelihood estimation problem~\eqref{eq:MLE} yielding new estimates for the line parameters $y$ and an updated covariance matrix $\Sigma_0$.
After running the adaption of $\rho$ from~\eqref{eq:rhoUpdate}, new inputs are computed by means of the combined OED-OPF problem \eqref{eq:oed}.
After applying these new inputs, the algorithm starts from the beginning.

%


\section{Numerical Case Study} \label{sec:numRes}

\begin{figure}
	\centering
	\includegraphics[width=0.75\linewidth]{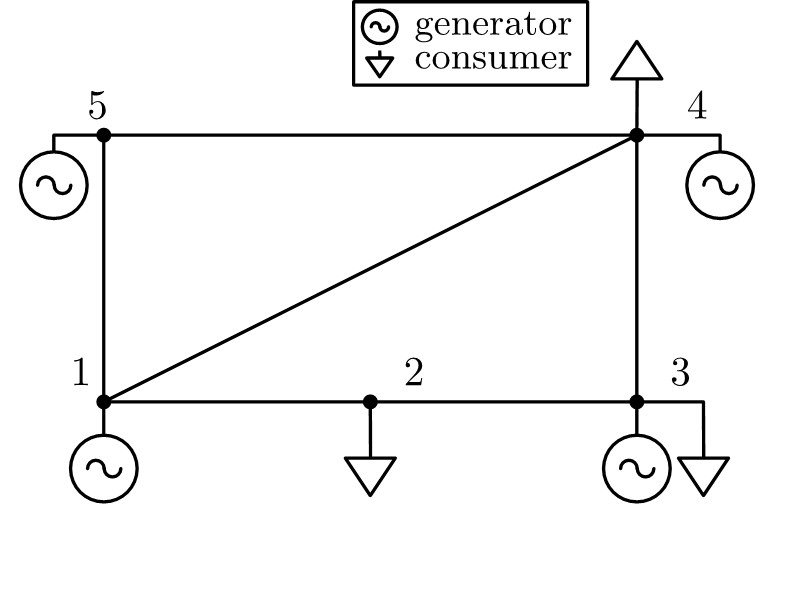}
	\caption{ Modiﬁed 5-bus system from Li and Bo (2010) with 4 generators and 3 consumers.}
	\label{fig:5buswithoutsplitting}
\end{figure}

Next, we illustrate the performance of Algorithm \ref{alg:OED} on a 5-bus system shown in  \autoref{fig:5buswithoutsplitting}.

\subsection{Implementation and Data}
The problem data is obtained from the \texttt{MATPOWER } \cite{Zimmerman2011}, where we neglect shunt capacities.
The implementation of Algorithm \ref{alg:OED} relies on \texttt{Casadi-v3.4.5} with \texttt{IPOPT} and \texttt{MATLAB 2019b}.
The cost  coefficients for the OPF cost $C$ from \eqref{eq:OPF2} are given in \autoref{table:result}.

\begin{table}[t]
	\centering
	\renewcommand{\arraystretch}{1.3}		
	\caption{Generator cost coefficients for \eqref{eq:OPF2}. } 
	\begin{tabular}{r||p{1.4cm}p{1.4cm}} 
		\hline
		 Bus Number&$\alpha_i$&$\beta_i$ \\
		\hline
	 	 1&0.1& 15 \\ 
		 3&0.11 & 30\\
		 4&0.12 & 40\\ 
		 5&0.13& 10\\
		\hline
	\end{tabular}
	\label{table:result}
\end{table}

We use additive white Gaussian measurement noise with zero mean and  variance $10^{-4}$, which is frequently considered used in context of power system parameter estimation \cite{Wood2013}.
We choose $\rho^0 = 10^{-4}$ for initialization and a sampling time of $15\text{min}$.
We initialize the parameters $y$ with the average true values of the admittance and the initial covariance matrix is set to $\Sigma_0=10^{20}\cdot I\in \mathbb{R}^{2|\mathcal{L}|\times 2|\mathcal{L}|}$.  
Moreover, we set $\mathrm{Tr}(\mathbb{V}_N^f)=10^{0} \,\mathrm S^2$ as the target variance after $25$ iterations.

%

\autoref{fig:V-rho} shows the dependency of $\frac{1}{\mathrm{ Tr}(\mathbb{V})}$ on $\rho$ for our numerical example after the first iteration in blue.
Moreover, \autoref{fig:Pareto} shows the corresponding Pareto-optimal curve of problem~\eqref{eq:OED_OPF} for different values of $\rho$, where we used the Pareto filter\footnote{In multi-objective optimization problems, the optimal solution is usually not a single one but a set of local optimal solutions of non-convex problems. A Pareto filter is used to filter out partial local optimal solutions to obtain monotonically decreasing Pareto front.} from \cite{Logist2012} to remove  local optima. 
We fit an  exponential function
\[\varphi\left (\frac{1}{\mathbb{V}}\right ) \approx 0.05891\exp^{-1411(\mathbb{V})^2}\]
for the usage in Algorithm~\ref{alg:OED} shown in \autoref{fig:V-rho} (red).


\begin{figure*}
	\centering
	\begin{subfigure}[t]{.47\textwidth}
		\centering
		\includegraphics[width=0.9\linewidth,trim=0 0 0 0, clip]{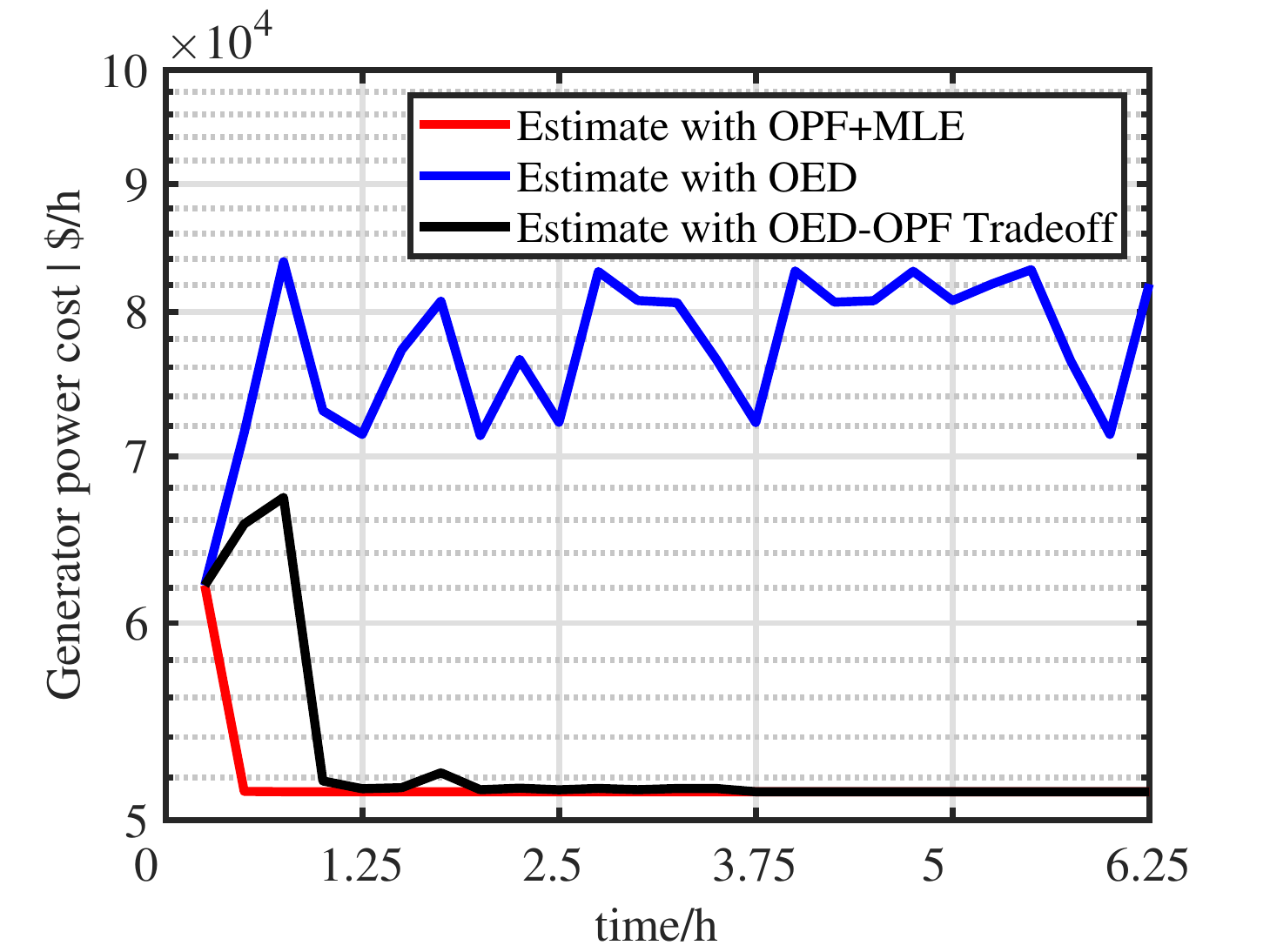}
		\caption{Operation cost of these three methods. }
		\label{fig:cost3}
	\end{subfigure}
	\vspace{1mm}
	\begin{subfigure}[t]{.47\textwidth}
		\centering
		\includegraphics[width=0.9\linewidth,trim=0 0 0 0, clip]{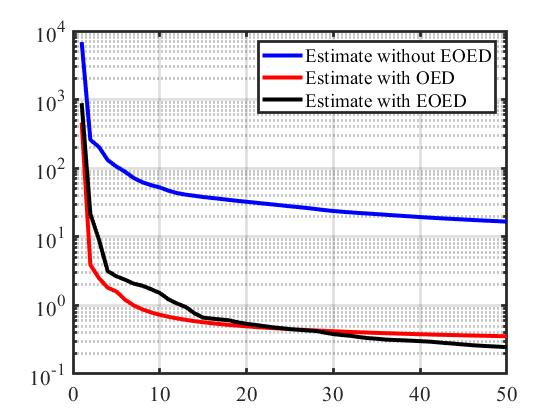}
		\caption{Variance of the line parameters $y$. } 
		\label{fig:variance}
	\end{subfigure}
	\begin{subfigure}[t]{.47\textwidth}
		\centering
		\includegraphics[width=0.9\linewidth,trim=0 0 0 100]{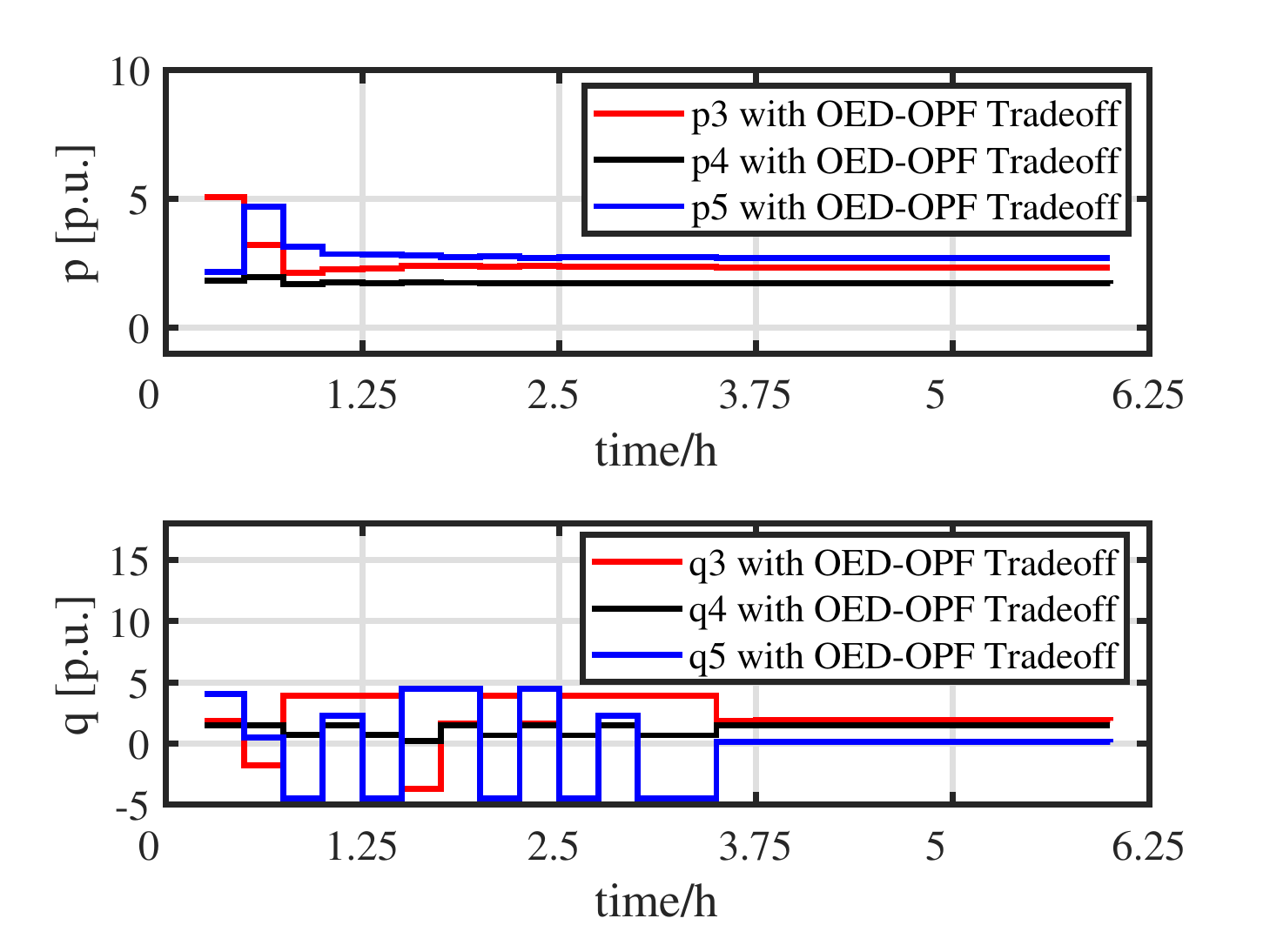}
		\caption{Optimal active and reactive power injection from Algorithm~\ref{alg:OED}. }
		\label{fig:EOED}
	\end{subfigure}
	\begin{subfigure}[t]{.47\textwidth}
		\centering
		\includegraphics[width=0.9\linewidth]{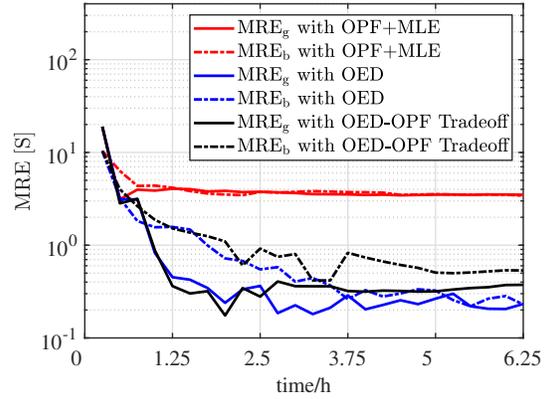}
		\caption{Mean relative errors of the line parameters $g$ and $b$. }
		\label{fig:ERROE}
	\end{subfigure}
	\caption{Estimation performance  and operation cost for all considered estimation algorithms.}
	\label{fig:compareAll}
\end{figure*}

\subsection{Numerical Comparison }

Next, we  compare the performance of Algorithm~\ref{alg:OED} to pure OPF combined with maximum likelihood estimation and pure OED in terms of operation cost and variance of the line parameters.

\autoref{fig:compareAll} shows the estimation performance and associated costs for all algorithms.
With a desired target variance of $\mathrm{Tr}(\mathbb{V}_N^f)=10^{0} \,\mathrm S^2$, 
Algorithm~\ref{alg:OED} leads to a substantially reduced cost compared with classical OED, see \autoref{fig:cost3}.
Classical OPF with MLE on the other hand is agnostic to the estimation variance and optimizes only with respect to the associated cost.
Hence, classical OPF with MLE leads to a slightly improved cost compared to Algorithm~\ref{alg:OED} but it comes with the disadvantage of a significantly worse estimation performance, see \autoref{fig:variance}.
The desired target variance is reached after $6.26$ hours which is only $25$ iterations with Algorithm~\ref{alg:OED}.

\autoref{fig:EOED} depicts the optimal active and reactive power input of the generators for Algorithm~\ref{alg:OED}. 
One can see that the active power is changed only in the first few steps and stays almost constant after that. The reactive power varies for more iterations.
Pure OED, however, leads to very frequent set-point adjustments in reactive \emph{and} active power, since it is agnostic to the associated economic cost  \cite{Du2020}.
This behavior is beneficial from a practical perspective: in contrast to changing the active power set-points, changing the reactive power set points  it is technically much simpler and cheaper.


\autoref{fig:ERROE} shows the mean relative errors
\begin{align*}
\mathrm{MRE}_\mathrm{g}=\frac{1}{|\mathcal{L}|}\sum_{(k,l)\in \mathcal{L}}\frac{|g_{k,l}-\bar g_{k,l}|}{|\bar g_{k,l}|} \;,\\
\mathrm{MRE}_\mathrm{b}=\frac{1}{|\mathcal{L}|}\sum_{(k,l)\in \mathcal{L}}\frac{|b_{k,l}-\bar b_{k,l}|}{|\bar b_{k,l}|}\;,
\end{align*}
in the estimation to $g_{k,l}$ and $b_{k,l}$  for all three methods.

\begin{table}
	\centering
	\renewcommand{\arraystretch}{1.3}		
	\caption{Estimation results for Algorithm~\ref{alg:OED}.} 
	\begin{tabular}{r||p{1.4cm}p{1.4cm}p{1.4cm}p{1.4cm}} 
		\hline
		Line &Conductance & Conductance  & Susceptance &Susceptance \\Index
		& true val. $[\mathrm{S}]$ & estimate $[\mathrm{S}]$ & true val. $[\mathrm{S}]$ & estimate $[\mathrm{S}]$ \\
		\hline
		$(1,2)$&3.523 & 3.529 &-35.235&-35.322 \\ 
		$(1,4)$&3.257 & 3.167& -32.569&-32.703\\
		$(1,5)$&15.470 &14.445&-154.703&-153.152\\
		$(2,3)$&9.168 & 9.429&-91.676&-91.746\\
		$(3,4)$&3.334 & 3.896&-33.337 &-32.474\\
		$(4,5)$&3.334 &3.223& -33.337  &-33.305\\
		\hline
	\end{tabular}
	\label{table:result1}
\end{table}

Table~\ref{table:result1} shows the ground truth $\overline{y}$ and the OED estimation result after 25 iterations. One can see that in all cases the relative error is below $10\%$.
\section{Summary and Outlook}
This paper presented a parameter estimation method for simultaneously minimizing operation cost based on optimal power flow and estimating line parameters.
An example shows that Algorithm~\ref{alg:OED} achieves a higher estimation accuracy compared with classical estimation methods and at the same time it is cheaper than strategies purely based on the optimal design of experiments.

Future work will consider advanced weighting strategies aiming at replacing the offline scheme adopted in this article. A numerical comparison between multi-stage OED-OPF problem and the penalty-function modified single-stage model will also be discussed.
\bibliographystyle{unsrt}
\bibliography{paper,alex}
\balance
 
\end{document}